\documentclass[11pt,twoside]{article}
\usepackage{asp2010}

\resetcounters
\def\kms{km~s$^{-1}$}

\begin{document}

\title{Galactic Small Scale Structure Revealed by the GALFA-HI Survey}
\author{Ayesha Begum$^1$,  Sne\v{z}ana Stanimirovi\'{c}$^1$, Joshua E.
Peek$^2$, Nicholas, Ballering$^1$,
Carl Heiles$^3$, Kevin A. Douglas$^4$,
Mary Putman$^2$, Steven Gibson$^5$, Jana Grcevich$^2$, Eric 
Korpela$^6$, Min-Young Lee$^1$, Destry Saul$^2$, Jay Gallagher$^1$
\affil{$^1$University of Wisconsin, Madison, 475 N Charter St, Madison, WI 53703}
\affil{$^2$Department of Astronomy, Columbia University, New York, NY 10027, USA}
\affil{$^3$Radio Astronomy Lab, UC Berkeley, 601 Campbell Hall, Berkeley, CA 94720}
\affil{$^4$ School of Physics, University of Exeter, Stocker Road, Exeter,
UK EX44QL}
\affil{$^5$ Department of Physics and Astronomy, Western Kentucky University,
Bowling Green, KY 42101}
\affil{$^6$ Space Sciences Laboratory, University of California, Berkeley,
CA 94720}}

\begin{abstract}

The Galactic Arecibo L-band Feed Array HI (GALFA-HI) survey
is mapping the entire Arecibo sky at 21-cm, over a
 velocity range of $-$700 to +700 km s$^{-1}$ (LSR), at a
  velocity resolution of 0.18 km s$^{-1}$ and an angular
resolution of 3.5 arcmin.
The unprecedented resolution and sensitivity of the GALFA-HI survey
have resulted in
the detection of  many~isolated, very compact HI clouds at low Galactic
velocities which are distinctly separated from the HI disk emission.
In the limited area
of $\sim$4600 deg$^2$ searched so far, we have detected 96 such compact clouds. The
detected clouds are  cold with T$_k$ less than 300 K. 
Moreover, they   are quite compact and faint,
with median values of 5 arcmin in angular size, 0.75 K in peak brightness
temperature, and  $5\times10^{18}$ cm$^{-2}$ in HI column density. 
From  the modeling of spatial and velocity distributions of the whole compact
cloud population, we find that the bulk of  clouds are related
to the Galactic disk, and are within a few kpc distance.
We present  properties of the compact clouds sample and discuss 
various possible scenarios for the origin of this clouds population
 and its role in the Galactic interstellar medium studies.
\end{abstract}

\section{Introduction}

Traditionally, HI observations have been able to trace
the entire hierarchy of structures in the diffuse interstellar
medium (ISM)  on scales $\geq$ 1 pc.  However, the small-scale end
of this spectrum, i.e scales $<$ 1 pc, is still largely unexplored because
of a paucity of high spatial/velocity resolution imaging surveys.
The presence of sub-pc scale ISM clouds  raises many interesting questions.
For example, how abundant is this cloud population?
What are the formation and survival mechanisms for sub-pc clouds with low
HI column densities? Also, what role do these clouds play in the general
ISM?  The small scale structure in the ISM is particularly interesting as it tends
to probe dynamic ``events" such as 
stellar winds \citep{matthews_2008,gerard_2006},
shocks \citep{gibson_2005}, turbulence \citep{avillez_2005,audit_2005} and
Galactic accretion \citep{heitsch_2009}. To fully investigate the nature
of small scale ISM clouds a sensitive, unbiased, high resolution survey of the
entire sky is required.

The Galactic Arecibo L-band Feed Array HI (GALFA-HI) survey is
successfully mapping the entire Arecibo sky at 21 cm. The survey
covers a velocity range of $-$700 to +700 km s$^{-1}$ (LSR)
at an unprecedented velocity resolution of 0.18 km s$^{-1}$ and
a angular resolution of 3.5 arcmin \citep{snez_2006,peek_2008}. The combination of sensitivity
and resolution provided by the GALFA-HI survey allows us to probe a new
regime of faint, small HI objects that have not been seen before
in lower resolution survey, e.g. Leiden/Argentine/Bonn (LAB) survey, 
Galactic all sky survey (GASS) \citep{gass_2009,lab_2005} or lower 
sensitivity surveys, e.g. Canadian Galactic Plane Survey
\citep{stil_2006}.
In this proceedings we present  properties of compact clouds
detected at Galactic velocities in the GALFA-HI survey and discuss 
their origin and role in Galactic ISM studies.

\section{Compact clouds search}

\begin{figure*}
\epsscale{1.0}
\plottwo{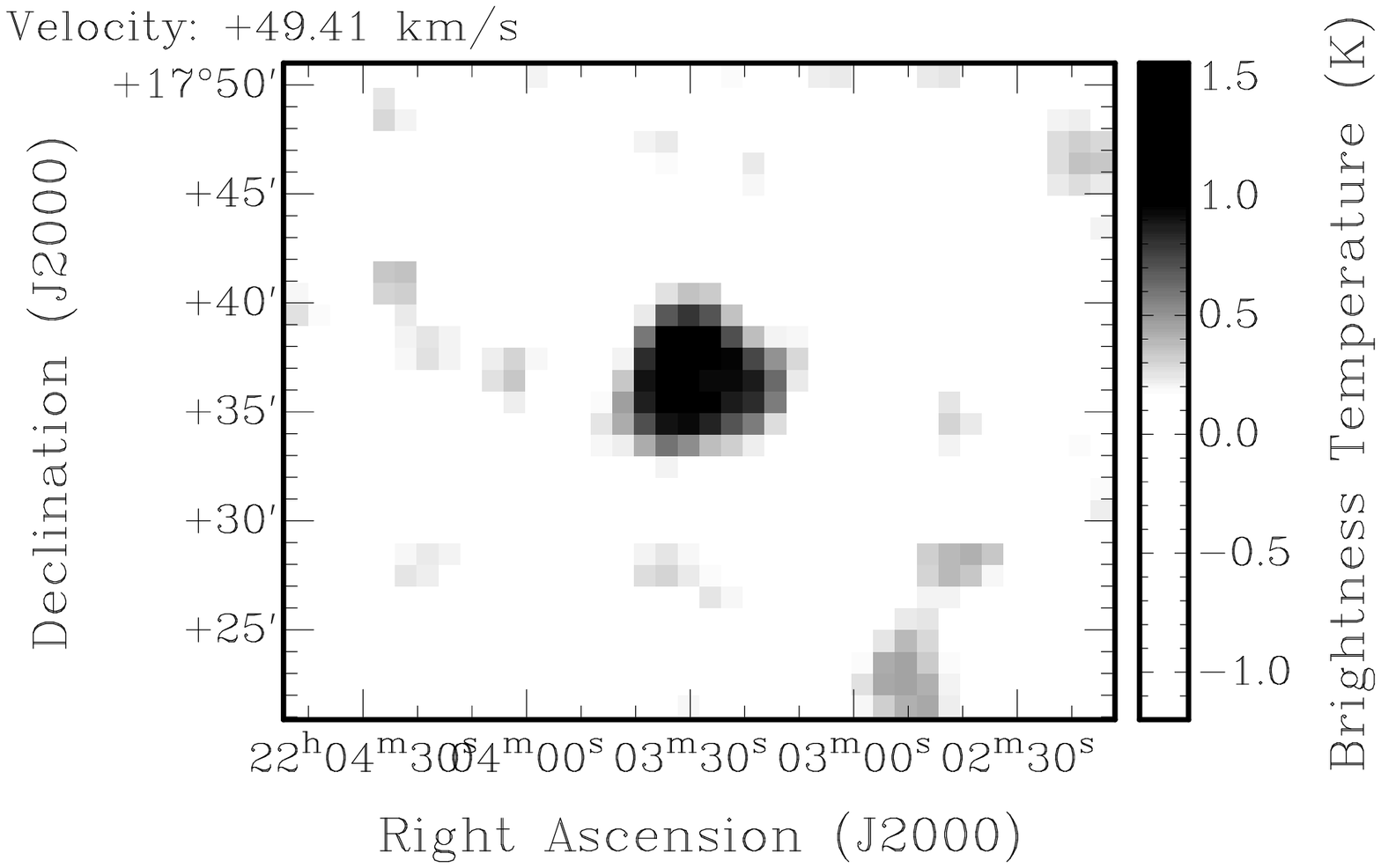}{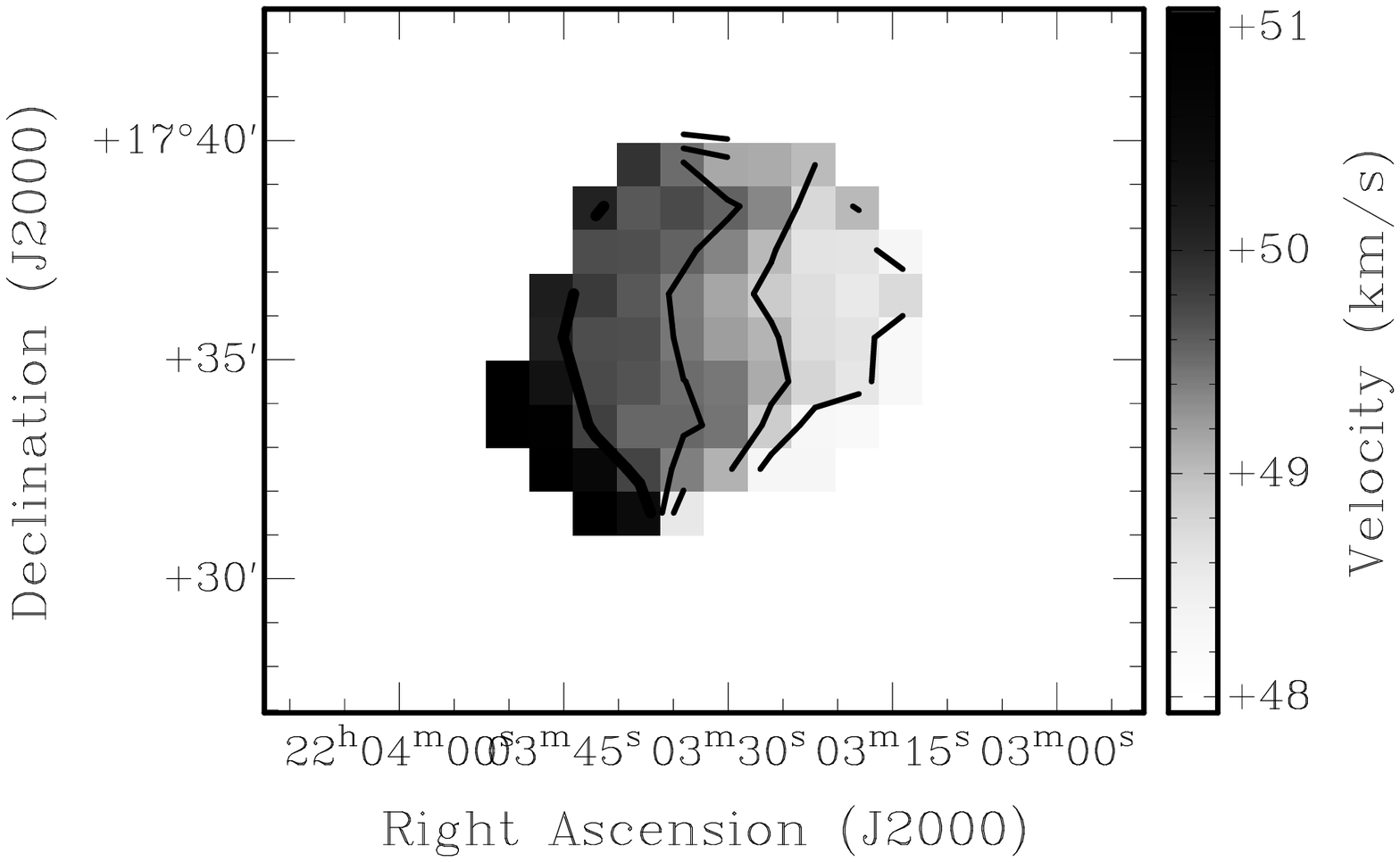}
\caption{\label{f:image1} 
{(\it Left)} The GALFA-HI image of a compact cloud at V$_{LSR}\sim 49$ \kms.
{(\it Right)} The velocity filed of the compact cloud.
}
\end{figure*}

The GALFA-HI data cubes were searched for {\it compact} (``almost unresolved")
 and {\it isolated} clouds (distinguishable from, if any, unrelated 
background emission). The  clouds 
which appeared to be a part of  larger,  filamentary 
structures were not considered. We 
searched for and identified clouds visually and  mainly found within 
the velocity range of $-120.0~<~V_{LSR}~<120.0$ \kms.
Compact clouds found outside this range were relatively rare and were
mainly identified as known galaxies. For details please see \cite{begum_2010}.
In the limited area of $\sim$ 4600 deg$^2$ searched so far, a total of 96 clouds were identified. 
An example image at a velocity of $\sim 49.0$ \kms, showing a compact HI
clouds, is presented in Figure~\ref{f:image1}(Left). The velocity field of 
this cloud is shown in Figure~\ref{f:image1}(Right).
The cloud properties are presented and discussed in the following sections.

\section{Properties of compact clouds}

Figure~\ref{f:hist}  show the histogram of some of 
the observed properties of the compact cloud sample.
We summarize the main results below.

\begin{itemize}

\item As shown in Fig.~2a, the clouds are typically compact, with the median angular size of
the sample being $\sim 5'$. They are either unresolved at Arecibo's
resolution or have  an unresolved core along  with some faint diffuse 
emission.

\item Fig.~2c shows that the majority of clouds have $T^{pk}_b=0.5-2$ K. The median peak
brightness temperature for the whole sample is 0.75 K.
This low peak brightness temperature, coupled with the small angular
size, explains why such clouds were largely missed by previous
large-scale Galactic HI surveys, e.g. the LAB and GASS survey \citep{gass_2009,lab_2005}.

\begin{figure*}
\epsscale{0.5}
\plotone{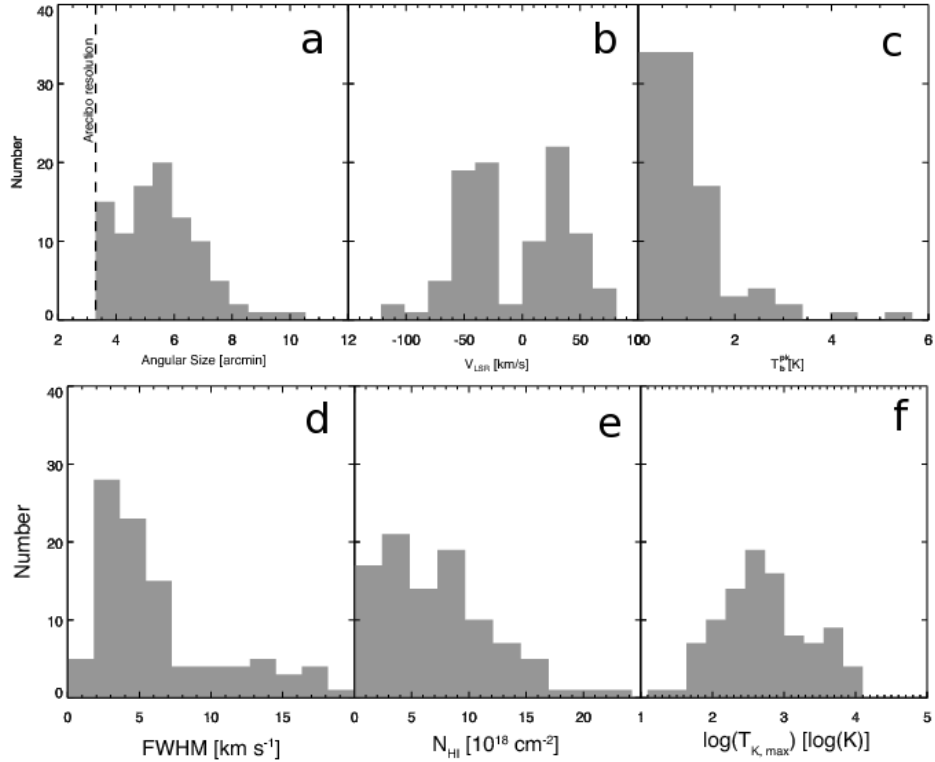}
\caption{\label{f:hist} Histograms of basic observed properties for
the whole sample of compact clouds: angular size, central V$_{LSR}$, the peak
brightness temperature, T$_b^{pk}$, FWHM, N$_{HI}$ and log(Kinetic temperature)
for the compact cloud sample.
}
\end{figure*}

\item Clouds are found at both positive and negative LSR velocities (see Fig.~2b). The
$V_{\rm LSR}$ histogram has a nearly symmetric distribution around
$V_{\rm LSR}=0$ \kms.
The gap  at  velocities between $-20 \leq V_{\rm LSR} \leq 5$ km~s$^{-1}$ is
not real, but due to difficulties in finding clouds in the presence of
bright Galactic emission.

\item Most  clouds are found in the first and second Galactic Quadrant, at least
partially due to our  limited survey coverage and appear at velocities both
allowed and forbidden by the Galactic rotation. We find that~most of the clouds
deviate from Galactic rotation  by at-most 50 kms$^{-1}$ level,
 i.e. deviation velocity $V_{\rm dev} \leq 50$  \kms~
(with a majority showing $V_{\rm dev} \sim$ 20 km~s$^{-1}$). 
Further, most of the clouds are at high Galactic latitudes with 
$|b| \sim 20^\circ-60^\circ$. 

\item The compact cloud sample has a narrow velocity line widths, in the range of $\sim$ 1$-$8
\kms,  with a median FWHM of 4.2 \kms (see Fig.~2d).  This corresponds to T$_{k,max}$$\sim$300 K,  
the upper limit on the kinetic temperature (i.e.~the kinetic temperature
in the case of no non-thermal broadening, defined as T$_{K, max}=21.86
\Delta V^2$). Our median
$T_{\rm k, max}$ is very similar to what is
found for the Galactic CNM clouds seen in absorption in the 
Millennium survey (Heiles \& Troland 2003), suggesting that these compact 
clouds have properties similar to those of typical Galactic CNM clouds.

\item As can be seen in Fig.2e, the integrated HI column density distribution peaks at $5 \times
10^{18}$ cm$^{-2}$, with $\sim$ 80\%  of clouds having
N$_{HI}<10^{19}$ cm$^{-2}$. This is low relative to typical Galactic
CNM clouds, which tend to have N$_{\rm HI} \sim 10^{20}$ cm$^{-2}$ \citep{heiles_2003}.

\item  For $\sim$31\% of the compact clouds in our sample, two Gaussian 
functions were required to fit the observed velocity profiles.  A narrow and bright
Gaussian function is required to fit the line center, while a faint
and broad component is  needed to fit the line wings. This indicates the
presence of multi-phase medium in one third of the sample clouds.

\item 

Nearly 32\% of the compact  clouds in our sample show
large scale velocity gradients of $\sim 0.5 - 1$ km~s$^{-1}$~arcmin$^{-1}$.
This could be a signature of rotation in case of self-gravitating clouds.
We can test the hypothesis that clouds are gravitationally bound by comparing
their virial and  the HI mass.
The median M${\rm{_{vir} \times D^{-1}}}$ 
for the sample is $\sim$ 2 $\times 10^3$ M$_{\odot}$ kpc$^{-1}$,
whereas the median M$_{\rm HI} \times {\rm{D}}^{-2}$  for 
the sample is $\sim$ 0.1 M$_{\odot}$ kpc$^{-2}$, where D is the unknown cloud distance. 
Thus,
$\rm{ \frac{M_{vir}}{M_{HI}} \sim \frac{2 \times 10^4}{D(kpc)}}$. 
Hence, for  any reasonable distance, gravity is totally 
negligible, unless a huge amount of dark matter is invoked to stabilize compact clouds.
Other possibilities, such as velocity gradients being due to  cloud expansion
from   stellar mass loss  are more appropriate \citep{gerard_2006,matthews_2008}.


\end{itemize}

\section{Distance constraints}

\begin{figure*}
\begin{center}
\includegraphics[scale=0.4,angle=0]{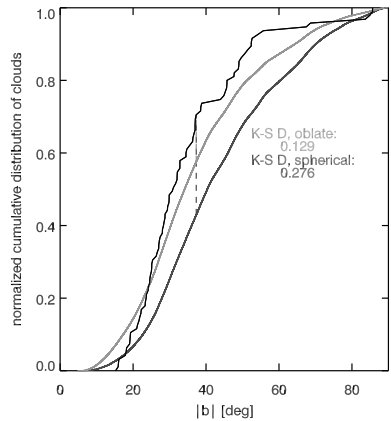}
\caption{\label{f:KSD}The normalized cumulative distribution of sample
clouds plotted as a function of Galactic latitude. The same quantity
is plotted for the two models with clouds having spherical
and oblate distributions. For details refer to \cite{begum_2010}.
}
\end{center}
\end{figure*}

In order to determine the nature of the compact clouds, we need to know their distance.
Are these clouds at extra-galactic distances or are they connected to the Galactic disk ?
One way we can check whether the compact clouds are connected to the Galactic disk is
by modeling their latitude distribution. A distribution which is oblate in the plane of the
Galaxy would indicate a relationship to the disk. Figure~\ref{f:KSD} shows the normalized cumulative
distribution of the compact clouds plotted as a function of the Galactic latitude. The same
quantity is also plotted for the two models with clouds having spherical and oblate 
distributions. As can be seen, a spherical distribution is a much poorer
fit to the data than an oblate distribution.
This suggests that the bulk of the clouds are related to the Galactic disk. 

Another constraint on the distance comes from a symmetric
distribution of  $\rm V_{LSR}$ around 0 kms$^{-1}$ (see Figure~\ref{f:hist}b). 
As most clouds in the sample
have $|b|>30$ degrees, a cloud distance $>3$ kpc would imply a height above the
Galactic plane that is $>1$ kpc. At such heights Galactic rotation is
lower than in the plane and this would result in a significant
lag to the cloud radial velocity \citep{levine_2008,collins_2002}.
As a result, the
 $\rm{V_{LSR}}$ distribution for the whole sample would be significantly skewed
toward negative velocities.
As we do not observe this effect when considering the whole sample
of compact clouds, we conclude that a majority of clouds are likely to be at a distance
$\leq3$ kpc.

\section{Discussion and conclusions}

As shown in the previous section, the bulk of the compact clouds are related to 
the Galactic disk and are within 2$-$3 kpc. We can now address the question
about what is the nature of compact clouds.

For a distance $\leq1$ kpc, a typical size of  compact clouds is $\leq1$ pc
and we are dealing with nearby Galactic clouds.
In this case,  a significant fraction of clouds would have
n$\geq$ 1 cm$^{-3}$,  a total pressure $\sim$ 1000 K cm$^{-3}$ and 
an HI mass $<10^{-1}$ M$_{\odot}$.
However, at a distance as low as 100 pc these properties become
slightly more extreme:
n$\sim$ 10 cm$^{-3}$, P$_{\rm ther} \sim$ 3000 K cm$^{-3}$,
$\rm{M_{HI}}\sim10^{-3}$ M$_{\odot}$, and
clouds would have a linear size of $\leq30,000$ AU.
In this scenario, the compact clouds probably
represent low column-density examples of the population of
CNM clouds on sub-pc scales.
The obvious questions are, how do such small and isolated clouds form
and survive in the ISM, and what role do they play within the general ISM?

One possible scenario that can explain the formation and maintenance of
compact cold clouds in the ISM is interstellar turbulence.
Recently numerical simulations of the ISM have started to describe cold
and warm atomic gas with a numerical resolution and dynamic range
approaching realistic physical scales e.g. \citep{avillez_2005,audit_2005}.
The numerical simulations by Audit \& Hennebelle (2005) show that a
collision of incoming  turbulent flows can initiate condensation of the WNM
into cold neutral clouds. A collision of incoming WNM streams creates a
thermally unstable region of higher density and pressure but lower temperature,
which further fragments into small  cool structures. The abundance of
cold structures,
as well as their properties, depends heavily on the properties of the underlying
turbulent flows. For example, \cite{audit_2005}
find that a significant fraction of the CNM structures formed in the case of
very turbulent flow have  thermal pressures of  $\leq 10^4$ K cm$^{-3}$,
temperatures $\geq$100 K and volume density n$\leq 100$ cm$^{-3}$. These CNM
structures are thermally stable, long lived and in the case of
stronger turbulence
they appear  round. These properties are similar to what we find for our
 compact cloud sample.

Another possibility for the formation of
compact clouds could be provided by stellar outflows.
HI emission has been detected in the circumstellar shells of a variety of
evolved stars, viz.~asymptotic giant branch stars, oxygen-rich and
carbon-rich stars,
semi-regular and Mira variables, and planetary nebulae \citep{gerard_2006,matthews_2008}.
Our cross-correlation of the compact cloud catalog with the
catalogue of variable stars by \cite{downes_2006}
suggests that a subset of clouds has at least one variable star within
a radius of 1 degree.
Assuming that the detected HI clouds are
the circumstellar HI associated with variable stars, for a typical
distance of 100 pc and an expansion velocity of 5 \kms~seen for circumstellar
HI \citep{gerard_2006},
a  one degree separation of the HI clouds from the variable star
corresponds to an HI diameter of 1.7 pc, a characteristic expansion time-scale
of 0.34 Myr and an HI mass of $\sim 2 \times 10^{-3}$ M$_\odot$. 
. All three parameters agree well with the ones found in
the HI survey of circumstellar
envelopes around evolved stars \citep{gerard_2006}. Hence it is likely
that some of the compact clouds are related to the outflows
from evolve stars.

Once formed by turbulence and/or stellar outflows and injected into the
surrounding  medium, these isolated compact clouds of cold, low column
density HI will be immersed in the warm/hot ambient gas. 
From Eq.~(47) in \citet{mckee}, the mass-loss of a compact HI
cloud embedded in a hot plasma ($T\sim 10^6$ K) is
$\sim 7 \times 10^{-2}$ M$_\odot$ Myr$^{-1}$, whereas if the clouds
are embedded in
large warm envelopes ($T\sim 7000$ K) the evaporation mass-loss is
ten times smaller.  This implies that the compact clouds may be
evaporating on a time-scale of $\sim$ 1 Myr, due
to a combination of conductive heat transfer and/or Kelvin-Helmholtz
instabilities
from the surrounding warm/hot medium \citep{snez_2005}.

So far we have considered the scenarios when the observed compact clouds are nearby i.e D $<$ 1 kpc.
However, if they are at a distance of a few kpc, the majority of clouds
would have  P$_{\rm ther} \leq$ 100 K cm$^{-3}$, n$\sim$ 1.0 cm$^{-3}$,
a height of $\sim$ 1 kpc above the disk, and corresponding HI masses of
$\sim 0.01 - 0.3$ M$_\odot$.
In this scenario, observed cloud properties are  similar to
numerous HI clouds found in the Galactic disk-halo interface region
\citep{lockman_2002,stil_2006,ford_2008,bekhti_2009}. 
The HI clouds in the disk-halo interface region are generally thought
to originate from the  condensation of hot gas expelled from the disk
by superbubbles \citep{hb90}.

To conclude, the unprecedented resolution and sensitivity of the GALFA-HI survey have
resulted in the detection of many compact, isolated, cold HI clouds at high Galactic 
latitudes \citep{begum_2010}. A significant fraction of 
these clouds show multi-phase medium and velocity gradients.
From the modeling and other distance constraints we 
find that the bulk of the clouds are likely to be related to the 
Galactic disk and are within a few kpc distance. Depending on the cloud distances, we have 
considered various possible scenarios for the
origin of this cloud population. If nearby at a distance less
than a kpc, these could be Galactic CNM clouds at sub-parsec scale,  formed
by stellar outflows and/or ISM turbulence. On the other hand, if the clouds are at a distance of
a few kpc, they are likely to be in the disk-halo interface region of the Galaxy.

We are currently developing automated methods for cloud detection.
Once the GALFA-HI survey is complete, we will collate a larger
sample of the compact clouds. This will allow us to quantify how
common such clouds are in the ISM, and whether they are related to
local events, such as stellar winds or
large scale atomic flows, or are globally distributed
across the disk with notable kinematic properties.

\bibliography{begum_ayesha}

\acknowledgements 
We are grateful to the staff at the Arecibo observatory for running 
the GALFA-HI observations. The Arecibo Observatory is part of the National
Astronomy and Ionosphere Center, which is operated by Cornell
University under a cooperative agreement with the NSF.
A.B., S.S., M.P., C.H., E.J.K., and J.E.P. acknowledge
support from NSF grants AST-0707597, 0917810,
0707679, and 0709347.

\end{document}